# Programmable Electric Tweezers

Yuang Chen[1], Haojing Tan[1], Jiahua Zhuang[1], Yang Xu[1], Chen Zhang[1,2]* and Jiandong Feng[1,2,3]*

[1]*Laboratory of Experimental Physical Biology, Department of Chemistry, Zhejiang University, 310027 Hangzhou, China*

[2]*The First Affiliated Hospital, School of Medicine, Zhejiang University, 310003 Hangzhou, China*

[3]*Institute of Fundamental and Transdisciplinary Research, Zhejiang University, 310058 Hangzhou, China*

*Correspondence to jiandong.feng@zju.edu.cn and chen-zhang@zju.edu.cn*

**Abstract**

The interaction mechanism between a single microscopic object like a cell, a particle, a molecule, or an atom and its interacting electromagnetic field is fundamental in single-object manipulation such as optical trap and magnetic trap. Function-on-demand, single-object manipulation relies on a high degree of freedom control of electromagnetic field at localized scales, which remains challenging. Here we propose a manipulation concept: programmable single-object manipulation, based on programming the electromagnetic field in a multi-bit electrode system. This concept is materialized on a Programmable Electric Tweezer (PET) with four individually addressed electrodes, marking a transition from function-fixed single-object manipulation to function-programmable single-object manipulation. By programming the localized electric field, our PET can provide various manipulation functions for achieving precise trapping, movement and rotation of multiscale single microscopic objects, including single proteins, nucleic acids, microparticles and bacteria. Implementing these functions, we are able not only to manipulate the object of interest on demand but also quantitatively measure the charge to mass ratio of a single microparticle via the Paul trap and the electrical properties of an individual bacterial cell by the rotation analysis. Finally, with superposed single-particle trapping and rotation, we demonstrate the spontaneous relaxation of DNA supercoiling and observe an unexpected pause phenomenon in the relaxation process, highlighting the versatility and the potential of PET in uncovering stochastic biophysical phenomena at the single-molecule level.

**Main:**

Past decades have witnessed the success of single microscopic object manipulation techniques such as optical tweezers and magnetic tweezers, in controlling the position of an atom[1], a molecule[2] or a cell[3]. These methodologies have enabled the exploration of the fundamental physical mechanisms ranging from quantum simulations to life phenomena at the ultimate limit[4-8]. Single-object manipulation, like trapping and rotation, is based on the interaction mechanism between a single object and its surrounding electromagnetic field. Therefore, the high degree of freedom in controlling the electromagnetic field is the core of function-on-demand, single-object manipulation. Existing techniques, such as magnetic tweezers[9-12] and optical tweezers[13-16], rely on the capability to control optical and magnetic field, whose functions are still highly fixed and remain hard to tune due to limited control dimensions. In contrast, localized electric field can be precisely manipulated by directly controlling the electric potential applied to the electrodes, setting the stage for single-object manipulation with electric means[17-19]. However, due to the limited degree of control over localized electric field in these systems, function-on-demand, single object manipulation remains an unaddressed matter.

Here we introduce programmable single-object manipulation, a concept based on programmable electric field control through multiple independent microscopic electrode systems. This approach overcomes the limitations of high degree of freedom control and can realize high-precision and high-flexibility generation of highly localized electromagnetic field patterns, to meet the needs of function-on-demand operation of multiscale single objects down to the nanometer scale. In this scenario, using an analogy to bit control in computing, the "electrode bit" refers to an individually addressable electrode and the voltage applied to the electrode corresponds to the information encoded within a bit.

Using this concept, we have developed programmable electric tweezers (PET). PET was equipped with four independently controlled microscopic carbon electrodes, which allow for the spatiotemporal programming of local electric field patterns to implement various electrokinetic mechanisms[20-24] for precise manipulation control. This spatiotemporal programming capability enables PET to selectively manipulate of multiscale single objects at the micro or nanoscale without affecting nearby microscopic objects. Specifically, our PET allows various manipulation functions including trapping, positioning and rotation. We further obtained the intrinsic properties such as charge to mass ratio and conductivity of the target object. The programmable function allows us to demonstrate superposed trapping and rotation, which enables the first measurement of the spontaneous relaxation process of intact DNA supercoiling process, a measurement that cannot be performed with optical tweezers and magnetic tweezers, thus allowing us to observe complex DNA mechanics at the single-molecule level.

## Results and Discussion

### PET concept

Our PET was equipped with four microscopic carbon electrodes fabricated from laser-pulled quartz capillaries[25-28] with adjustable gaps (**Supplementary Note 1**). The system was integrated into a three-dimensional micromanipulator **(Fig. 1a)** capable of high-precision localization and movement. By designing specific electrical commands ($S_{1A}$- $S_{4X}$) sent to each microscopic electrode (**Fig. 1b**), the voltage landscape can be spatiotemporally programmed within the highly localized region surrounded by quadruple microscale/nanoscale electrodes (**Fig. 1c**), resulting in multiple programmable functions ($F_A$- $F_X$) as shown in **Fig. 1d.** The frequency, the amplitude and the phase of these applied electrical signals on each electrode can be adjusted on-demand to tailor the localized electric field patterns, further utilizing multiple electrokinetic phenomena to achieve different functions. For example, trapping and rotation can be achieved by programming the phase of the signal on microscopic electrodes based on principles of Paul trap[22] and electrorotation (ROT)[20], as discussed in **Supplementary Note 2**. A higher degree of freedom manipulation requires more control dimensions, which can be realized by increasing the number of the 'electrode bit'. With an arrayed electrode system (**Extended Data Fig. 1**), arbitrary patterns of localized electric field (**Extended Data Fig. 2**) can be generated by programming each 'electrode bit', thereby realizing function-on-demand manipulation (**Supplementary Note 3**).

### Single-object trapping and measurement

As in optical tweezer experiments, we first verified the effectiveness of the trapping function of PET by trapping and releasing single 1 μm polystyrene (PS) beads (**Fig. 2a** and **Supplementary Video 1)**. Fluorescently labeled bead could be trapped at the center of the PET via the application of alternating current (a.c.) voltages. When the voltages were withdrawn, the trapped bead was released and then diffused away. The same process can be repeated for multiple cycles, demonstrating the high repeatability and the stability of PET. In addition, by scaling down the gap distance between microscopic carbon electrodes (8 μm to 400 nm tuning is demonstrated with our protocol, **Supplementary Fig. 3**), the PET can also trap nanoscale objects such as single proteins (**Extended Data Fig. 3 and Supplementary Video 2**) and nucleic acids (**Extended Data Fig. 4**), showing its capability to trap multiscale objects from single particles to single molecules (**Supplementary Note 4**).

The trapping stiffness $k = \frac{k_\mathrm{B} T}{\delta^2}$ can be obtained by fitting the transverse x-y position distributions of the trapped bead, where $k_\mathrm{B}$ is the Boltzmann constant, $T$ is the absolute temperature and $\delta$ is the standard deviation. We found the stiffness of the trapped beads to increase with the increase of voltages at a fixed frequency ($f$ = 3 MHz, **Fig. 2c**). Under a fixed voltage ($V_{ac}$ = 1.5 V), the influence of the frequency on the capture stiffness shows a non-monotonic trend[23], indicating the trapping stability is frequency-dependent (**Fig. 2d-e**). The trapping stiffness of single objects can be tuned by directly adjusting the voltage and the frequency to meet different experimental requirements across various objects. An additional advantage of our PET implemented on a glass capillary lies on its spatial positioning capability when a single object is stably trapped, which allows us to manipulate 1 μm PS beads to navigate a trajectory 'ZJU' (**Fig. 2f**) with 38 nm precision at 0.96 μm/s, demonstrating the nanoscale-controlled movement of a single target object.

Furthermore, the trapping stability provides a quantitative measurement of the intrinsic property of the object interacting with the electric field. Despite the presence of dielectrophoresis when time-varied electric field was applied, the dominant trapping mechanism in present configuration was determined to be Paul trap (**Supplementary Note 5** and **Supplementary Video 3**). Thus, PET can be used to measure the charge to mass ratio of a single object based on the trapping stability described by solutions of Mathieu equation[23]. Specifically, the stability of the trapped object in Paul trap is determined by parameters *a* and *q*, which correlates $U_{dc}$, $V_{ac}$ and frequency of signals with the charge to mass ratio of the object (detailed in **Supplementary Note 6**). By carefully adjusting $U_{dc}$ and $V_{ac}$ at a fixed frequency of 3 MHz to search for the verge where the trapping of bead was no longer stable, we obtained the boundary points, which were further plotted on the *a-q* diagram and compared with numerically calculated results (**Fig. 2g**), yielding an effective $\frac{Q}{\Gamma M}$ of 6.46×10$^{-5}$ e/amu of the target bead. By scaling down the size of the PET, multiscale objects, ranging from single particles to single molecules, may be potentially analyzed, opening a possibility for single-molecule mass spectrometry in solution.

**Single-object electrorotation**

Electrorotation (ROT)[20,29,30], a phenomenon that polarized objects rotate asynchronously with the rotating electric field, was adopted to realize the rotation function of PET. The electrorotation of partially fixed single bacterial cells (**Supplementary Note 7**) was investigated to demonstrate the rotation function of our PET and detailed voltage programing on PET electrodes is discussed in **Supplementary Note 8** and **Supplementary Video 4**.

**Fig. 3a** shows counterclockwise (CCW) rotation of a single *E. coli* cell (**Supplementary Video 5**) and intensity analysis in **Fig. 3b** reveals a constant time for one rotation revolution, corresponding to an angular rotation speed of 18 Hz (**Fig. 3c**). We then investigated the rotation direction and the angular rotation speed of *E. coli* at different voltages and frequencies. A linear correlation was observed between the angular velocity and the square of voltages at a fixed frequency (**Fig. 3d**), as quantitatively described in the model of electrorotation (**Supplementary Note 2**). Further, we observed a reversal of rotation direction of *E. coli* at various frequencies under a fixed voltage (**Fig. 3e**). The ROT spectra of *E. coli* were plotted by fitting the rotation rates with different frequencies as shown in **Fig. 3f**, from which, we obtained the physical properties of bacteria, such as the conductivity of the cytoplasm $\sigma_{cyto}$ = 4.45 mS·cm$^{-1}$ and the conductivity of the outer membrane $\sigma_{mem}$ = 31.6 μS·cm$^{-1}$ for *E. coli* (**Extended Data Fig. 5**). Meanwhile, *Bifidobacterium*, as a gram-positive bacterium, has a thick peptidoglycan cell wall rather than a lipid outer membrane in *E. coli* and a two-shell structure was adopted in its modeling (**Supplementary Note 9** and **Extended Data Fig. 6**). The different dielectric properties and structures may lead to the nonoccurrence of rotation reversal for *Bifidobacterium*, and the conductivity of cytoplasm $\sigma_{cyto}$ = 4.51 mS·cm$^{-1}$ as well as the conductivity of the outer cell wall $\sigma_{wall}$=

$1.76 \times 10^{-3}$ µS·cm$^{-1}$ were obtained. Because the conductivities of different kinds of bacteria are characteristic, we can distinguish different kinds of bacteria in situ by simply measuring the conductivity of single bacteria and can also measure conductivity heterogeneity in different periods of bacteria.

Compared with its chip-based counterparts[24,31-33], PET with a gap size of less than 10 µm can enhance the rotating electric field, boosting high-speed single object rotation. Single partially fixed *E. coli* on glass can be rotated up to a speed of ~5000 revolutions per minute. This capability of high-speed rotation is crucial for achieving measurements in special environments, such as low Reynolds number conditions[34,35]. In addition to rapid rotation, PET also features selective manipulation without affecting other objects (**Supplementary Video 6**). By combining these two key features: rapid rotation and selective manipulation, the PET system can achieve both measurements and precise manipulation, empowering the ability of PET to study targets in situ.

**Superposed multifunctional manipulation**

DNA supercoiling plays an important role in replication and transcription while double-strand breakage (DSB) can profoundly affect the stability of DNA supercoiling[36-39]. Since the supercoiling relaxation process induced by DSB is complex and instantaneous, a multi-step operation with superposed functions is required to simulate this process in single-molecule biophysics. Traditional optical tweezer and magnetic tweezer usually maintain constant supercoiling state and allow single-step mechanical loading, but remains difficult to implement a highly complex process such as "trapping-rotation-relaxation" due to the limitations in fast function switching and integration. As a result, important details of the relaxation dynamics following DSB remain elusive.

Our PET enables rapid switching and integration of different functions through real-time adjustments of programmable electric field. This facilitates multi-step manipulations and makes it possible to measure the relaxation dynamics after DSB. Here we used PET to program a multi-step manipulation sequence: trapping, trap with rotation, and back to trapping. We employed a configuration where a 1 μm bead was attached to 13 kbp torsional-constrained double-stranded DNA fixed on the glass, allowing rotational force transfer between the bead and the DNA (**Fig. 4a, Supplementary Note 7**). The application of the ROT signal caused the formation of DNA supercoiling, thus the decrease of Z position of the bead (**Supplementary Note 10**) can be observed (**Fig. 4a**). After removing the ROT signal, we observed a spontaneous relaxation process (~80 s to ~100 s). During this period, a rapid clockwise rotation as well as an increase in Z position of the bead was observed, implying the relaxation of DNA supercoiling. Interestingly, we observed a distinct pause in the last two supercoiling relaxations **(Fig. 4b)**, suggesting the existence of energy barriers during the spontaneous relaxation process of intact DNA supercoiling. In the rapid relaxation period, the elastic energy within DNA is enough to drive a fast relaxation without observable pauses. However, in the slow relaxation period, the supercoil elastic energy is reduced. The tensile force exerted by the PET and thermodynamic fluctuation may assist in overcoming the barrier, yielding the pause phases (**Fig. 4c**). Thus, we observed energy barriers during spontaneous DNA supercoiling relaxation, which has been predicted by the simulated DNA relaxation[40] but not directly been observed. Our finding provides a new experimental insight into to DNA supercoiling relaxation process, demonstrating that the high degree of freedom control in PET is beneficial to study transient and multi-step biophysical dynamics at the single-molecule level.

**Conclusion**

PET has demonstrated its versatile multiscale manipulation capabilities, enabling precise trapping, rotation, and measurement of single objects through spatiotemporally programmed electric fields. Among the new manipulation capabilities that PET can provide, we outline three main advantages owing to its 'multi-bit electrode' configuration and electric control nature: First, PET overcomes the limitations of traditional technologies such as optical tweezers and magnetic tweezers by enabling real-time function reshaping via programmable electric fields. This allows swift switching between multi-step functions, a critical feature for capturing transient biological processes. Second, PET features a scalable electrode gap size which controls the field strength, thus enabling direct manipulation of multiscale single objects across micro- to nanoscale range, from single bacteria and microparticles down to individual nucleic acids and proteins. Third, the integration of local electric field control within a glass micropipette confines manipulation only to the target area. Thus, selective manipulations of object of interest are highly feasible, with a potential to directly manipulate intracellular objects. These capabilities position PET as a transformative tool for biophysics, particularly for the measurement of transient dynamic processes at the single-molecule level. To demonstrate an example for practical application, we use

PET to observe the spontaneous DNA supercoiling relaxation. Moreover, PET has the potential to achieve more sophisticated function-on-demand manipulations by simply scaling up the number of electrode bits ($N$) as the allowed operations scale with $2^N$ (**Supplementary Table 4**), greatly enriching the function possibilities. Together, we believe PET has set a playground for high degree of freedom manipulation of single objects on demand, which may enrich the toolbox for addressing single entities, ranging from molecules, particles to cells, in both physical science and life science.

**Methods**

**Fabrication of PET.** PETs were fabricated from custom-designed quad-barrel quartz capillaries (manufactured by Zhong Cheng Quartz Glass) via a four-step process (detailed in **Supplementary Note 1**). Briefly, quad-barrel capillaries were first pulled using a two-line program via a laser puller (P-2000, Sutter Instrument). Next, butane was passed through the nanopipette and heated to deposit carbon electrode at the tip under a nitrogen atmosphere. The tip of nanopipettes was further etched by a 10:1 buffered oxide etchant solution (Sigma-Aldrich) to expose the carbon electrodes. As the final step, copper wires with one end soldered to a pin header were inserted into nanopipettes from the back end to establish the electrical connection, and a drop of glue (Ergo 5400) was added to complete the fixation and insulation of copper wires.

**Fluorescence imaging.** All fluorescence images and videos were acquired on an inverted optical microscope (IX83, Olympus) with a ×150 oil-immersion objective (1.45 numerical aperture, UApo N, Olympus) by an Electron Multiplying Charge-Coupled Device (EMCCD) camera (iXon Ultra 897, andor). Fluorescent PS beads with 1 μm size were illuminated by a 488 nm laser (C-FLEX, Hübner Photonics).

**Single particle manipulation and measurement.** PET was mounted on a mechanical micromanipulator or an electrical micromanipulator (in positioning experiment for a better position control, uMp, Sensapex) vertically with its pin headers connected to the Dupont cables which were linked to a function generator (AFG1062, Tektronix). In experiments, the PET was perpendicularly inserted to the sample cell which contains target particles (~1×10$^6$ particle/mL in ultrapure water) to obtain a clear imaging of PET and target single particles. After applying appropriate a.c. signals from the function generator, target particle can be captured when the PET approaches from above. Manipulations or measurements can be further performed by maneuvering the micromanipulator or changing parameters of a.c. signals, respectively.

**Electrorotation of single bacteria using PET.** In electrorotation (ROT) experiments, coverslips were treated with 0.1mg/mL Poly(L-lysine) (Adamas-life) to obtain partially fixed single bacteria. Bacteria were centrifuged and resuspended in ultrapure water at a proper concentration. The bright field imaging was acquired on a microscope (IX73, Olympus) with a ×100 oil-immersion objective (1.30 numerical aperture, UPlanFL N, Olympus) by a scientific Complementary Metal-Oxide-Semiconductor camera (Kinetix, Teledyne) for an exposure time (down to 2 ms) to capture the ultra-high rotation of bacteria. In experiments, PET was positioned by a mechanical micromanipulator to the location where single target was located at the center of PET. After applying phase-shifted signals from a function generator (AFG31000, Tektronix), target bacterium can be rotated and the rotation direction and speed can be regulated by adjusting the voltage, frequency and phase of signals.

**DNA supercoiling measurement.** Torsion-constrained DNA was incubated with anti-digoxigenin coated glass in 2 mL 3× PBS solution for 10 minutes, then half of solution was changed to 0.1× TE and streptavidin-modified PS beads or magnetic beads with 1μm size were added to a density of ~$5×10^5$ particles/mL. After incubation for 30 minutes, the solution was changed to ultrapure water for DNA supercoiling measurement. PET was vertically inserted into the solution to approach a DNA-linked bead. Custom-designed adders (AD812, Analog Devices) were connected to PET to apply super-positioned signals. The bead can be captured by applying Paul trap signal and further lifted by increasing the a.c. voltage. Then, the bead can be rotated with a defocus phenomenon, indicating the formation of DNA supercoiling if appropriate ROT signals were superposed. When the ROT signals were removed, the bead rotated in the opposite direction and returned to its initial trapped state eventually.

**Acknowledgments**

We thank Xi Zheng and Nianhang Rong at the Analysis Center of Agrobiology and Environmental Sciences for SEM characterization. This work was funded by the National Natural Science Foundation of China (grant number: 21974123, 12304254) and the National Key R&D Program of China (grant number: 2020YFA0211200). J.F. acknowledges the support from the New Cornerstone Science Foundation through the XPLORER Prize.

**Author Contributions**

J.F. conceptualized the project. J.F. and C.Z. supervised the project. J.F., C.Z., and Y.C. designed the experiments. Y.C. conducted the experiments. Y.C., H.T., Y.X. and C.Z. carried out the data analysis. J.Z. prepared the bacteria samples. Y.C. and Y.X. constructed the simulations. Y.C., C.Z. and J.F. wrote the manuscript.

**Data Availability**

Source data are provided with this paper. Data in other formats are available from the corresponding author upon request.

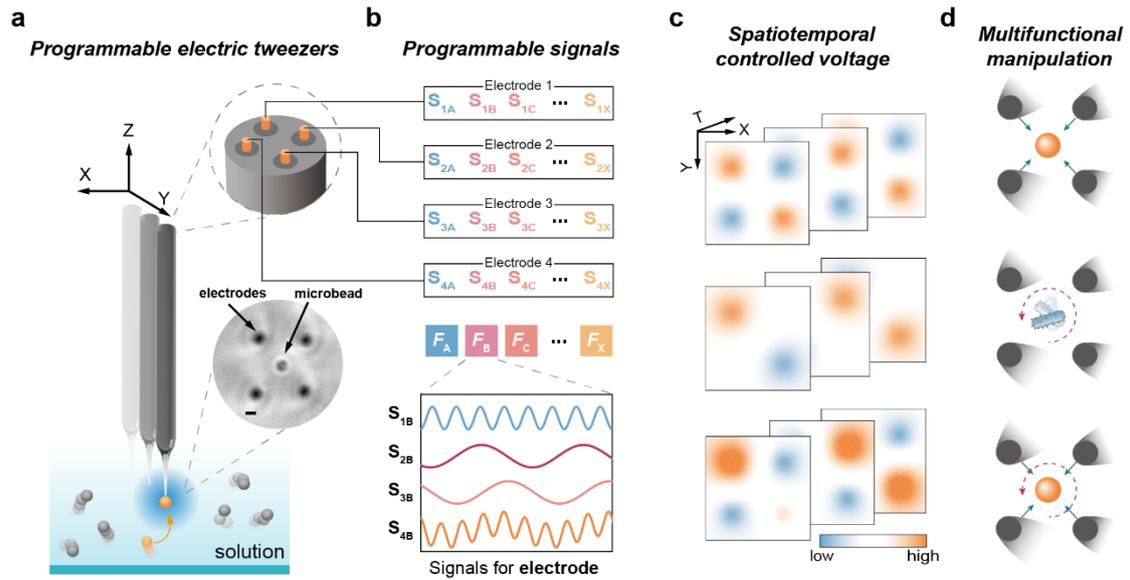

**Fig. 1 Principle for programmable electric tweezer (PET). a**, PET with four individual addressable electrodes was clamped vertically by micromanipulator and inserted into solution to achieve required manipulations and measurements. The inset bright field image shows single bead trapping by PET, scale bar: 1 μm. **b**, Programmed signals ($S_{A1}$, $S_{A2}$, …, $S_{X4}$) can be applied individually to realize multiple functions ($F_A$, $F_B$, …, $F_X$). numbers and letters in subscript represent electrodes and functions, respectively. **c**, Schematic for spatiotemporal controlled voltage patterns generated by PET. **d**, Schematic for corresponding multifunctional manipulations of single entities by spatiotemporal controlled voltage in **c**.

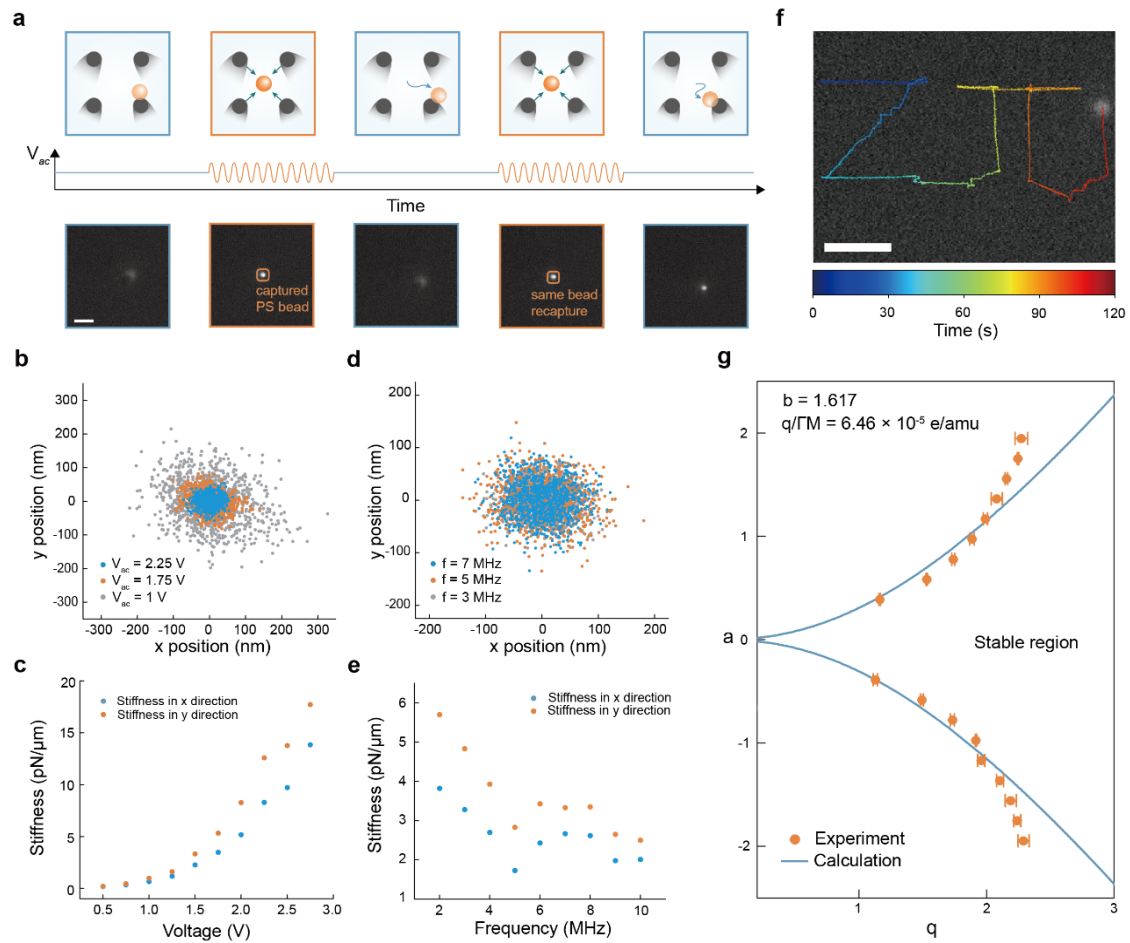

**Fig. 2 Single particle trapping, manipulation and measurement. a,** Capture, release and recapture of a 1 μm PS bead. Scale bar: 5 μm. **b**, Scatter plot of a trapped bead under typical voltages of $V_{ac}$ = 1 V (gray), 1.75 V (orange) and 2.25 V (blue) at a fixed frequency of 3 MHz. **c**, Stiffness for 1 μm PS bead trapping as a function of voltage (*f* = 3 MHz). **d**, Scatter plot of trapped bead at typical frequencies of *f* = 3 MHz (gray), 5 MHz (orange) and 7 MHz (blue) under a fixed voltage of $V_{ac}$ = 1.5 V. **e**, Stiffness for 1 μm PS bead trapping as a function of frequency ($V_{ac}$ = 1.5 V). **f**, 'ZJU' trajectory written by a trapped PS bead via the motion of quadruple electric tweezer ($V_{ac}$ = 1 V, *f* = 5 MHz). Scale bar: 5 μm. **g**, a-q stability diagram contains the boundary points obtained from experiment using $\frac{Q}{\Gamma M} = 6.46 \times 10^{-5}$ e/amu and curves obtained from the calculation

using $b$ = 1.617, which was detailed in **Supplementary Note 5**.

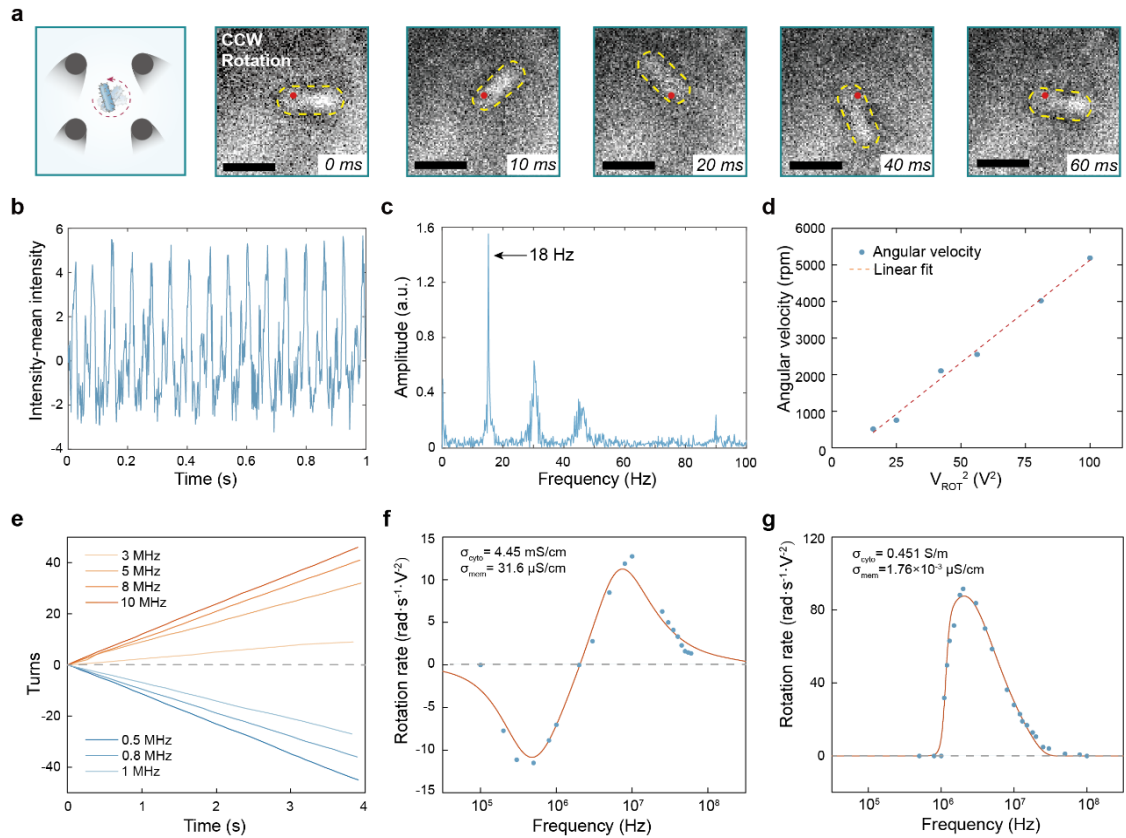

**Fig. 3 Rotation and measurement of single bacterium. a**, Schematic and snapshots for electrorotation of a single *E. coli*. Scale bar: 2 μm. Yellow dash line denotes the shape of *E. coli* and red point denotes the fixed part ($V_{ROT}$ = 5 V, $f$ = 20 MHz). **b**, Intensity-time trace showing a periodical change which corresponds to the rotation of *E. coli* with constant time intervals between each cycle. **c**, Fast Fourier transform (FFT) result of time trace in **Fig. 3b**. The highest peak stands for the real angular velocity of 18 Hz of rotation for *E. coli* while other peaks are multiple frequency peaks owing to the FFT process. **d**, Relationship between angular velocity and $V_{ROT}^2$ as described in the theory of electrorotation. **e**, Rotation speed (positive value represents CCW rotation) at different frequencies for *E. coli* under $V_{ROT}$ = 5 V. **f**, ROT spectra of *E. coli* and corresponding fitting. **g,** ROT spectra of *Bifidobacterium* and corresponding fitting.

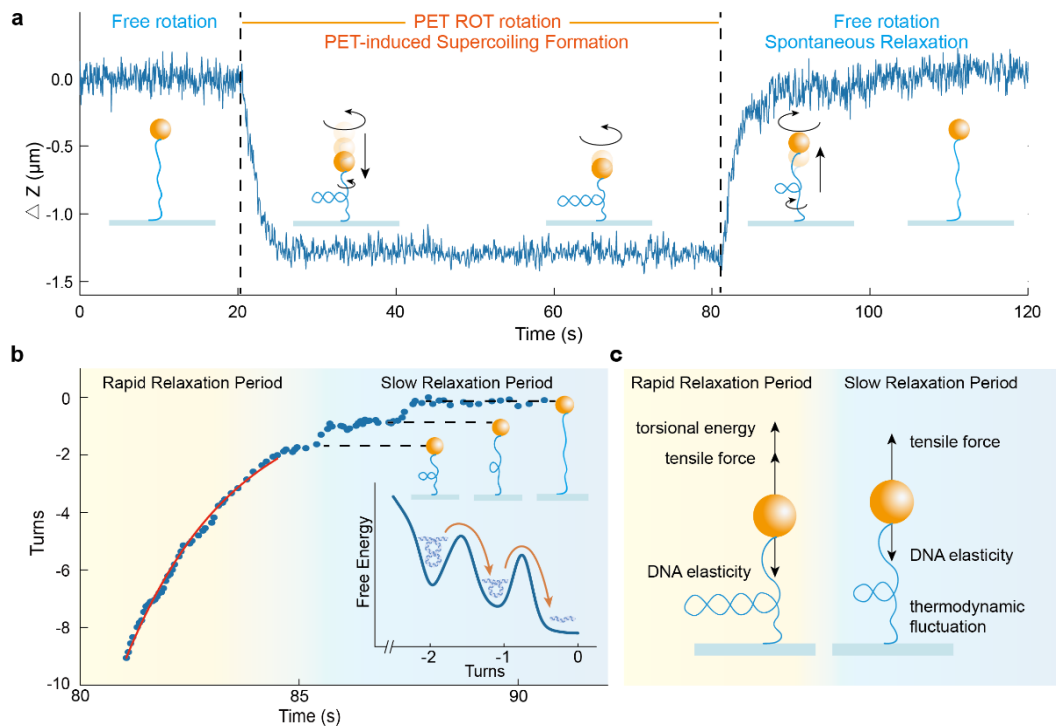

**Fig. 4 Multifunction manipulation for DNA supercoiling relaxation. a**, Displacement in Z direction versus time curve showing the formation and the relaxation process of DNA supercoiling. Lifted bead experiences a DNA supercoiling formation process (~ 20 s to ~ 80 s) with rapid and slight decreasing in Z position, respectively, indicating the introduction of plectonemes. When the ROT signals are withdrawn (~ 80 s), the bead experiences a spontaneous relaxation process with an increasing in Z position and finally returns to its original state ($V_{ac}$ = 2 V at $f_{ac}$ = 3 MHz, $V_{ROT}$ = 0.5 V at $f_{ROT}$ = 2 MHz). **b**, Time-turns sampling during the spontaneous relaxation process. Three pause phases are observed, implying nonlinear dynamics in spontaneous relaxation of DNA supercoiling. Inset is the schematic of energy diagram for the relaxation of last two supercoils. Multiple energy barriers are proposed along the relaxation process thus the overcoming of energy barrier (orange arrows) may be the combined effects of tensile force of Paul trap and thermodynamic fluctuation of DNA, yielding discrete plateaus. **c,** A schematic representation illustrating the forces exerted on microspheres during the two stages of DNA supercoiling relaxation.

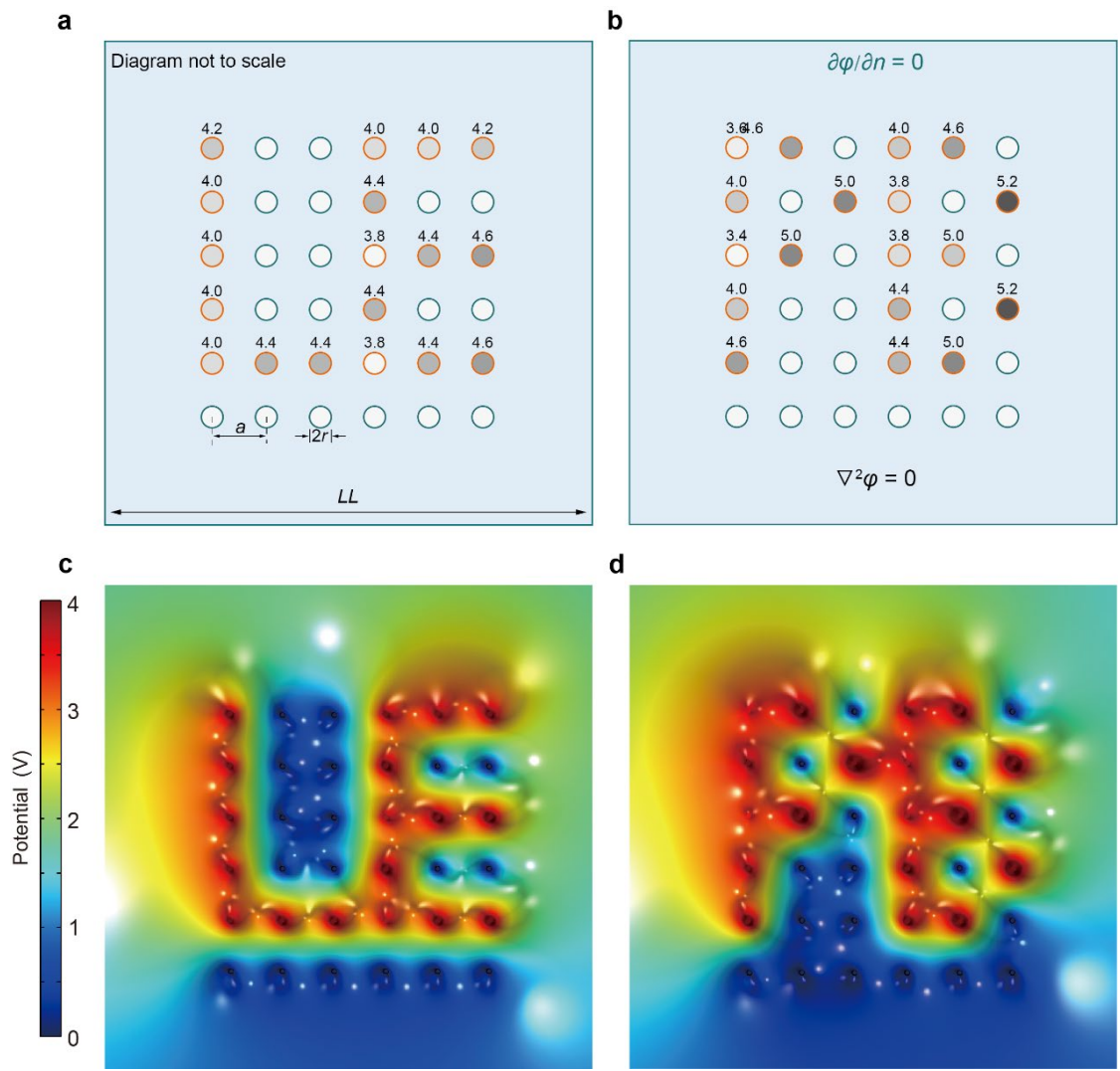

**Extended Data Fig. 1. Demonstration of local potential programmability of PET. a,** Geometry of the model and potential design for 'LE' pattern. The center of the 6×6 nanoelectrode array is located at the center of square. Numbers represent the voltages applied on each electrode with unit of volt and those electrodes without number and boundary of model were grounded. **b,** Boundary conditions of the model and potential design for 'PB' pattern. **c,** Potential distribution from the simulation results for 'LE' pattern. **d,** Potential distribution from the simulation results for 'PB' pattern.

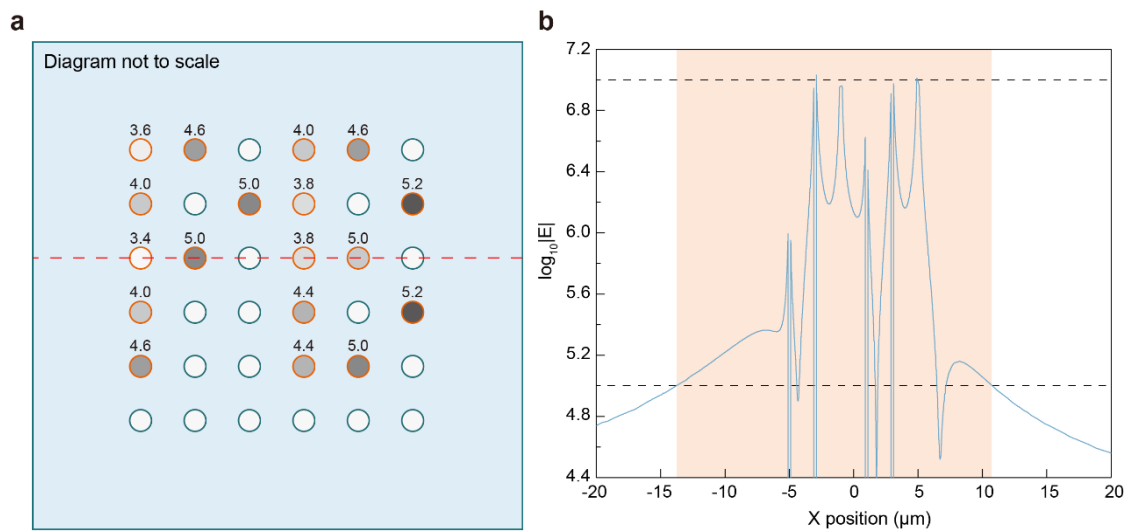

**Extended Data Fig. 2. Investigation in range influenced by electric field generated by PET. a**, The distribution of electric field strength along the red dashed line (across the center of 'PB' with the center point corresponding to X position = 0 in **b**) was extracted to investigate the range influenced by electric field. **b**, Electric field strength as a function of the x-position. Two dashed lines indicate the peak value and 1% of the peak value respectively and the orange square represents the region influenced by the electric field generated by PET.

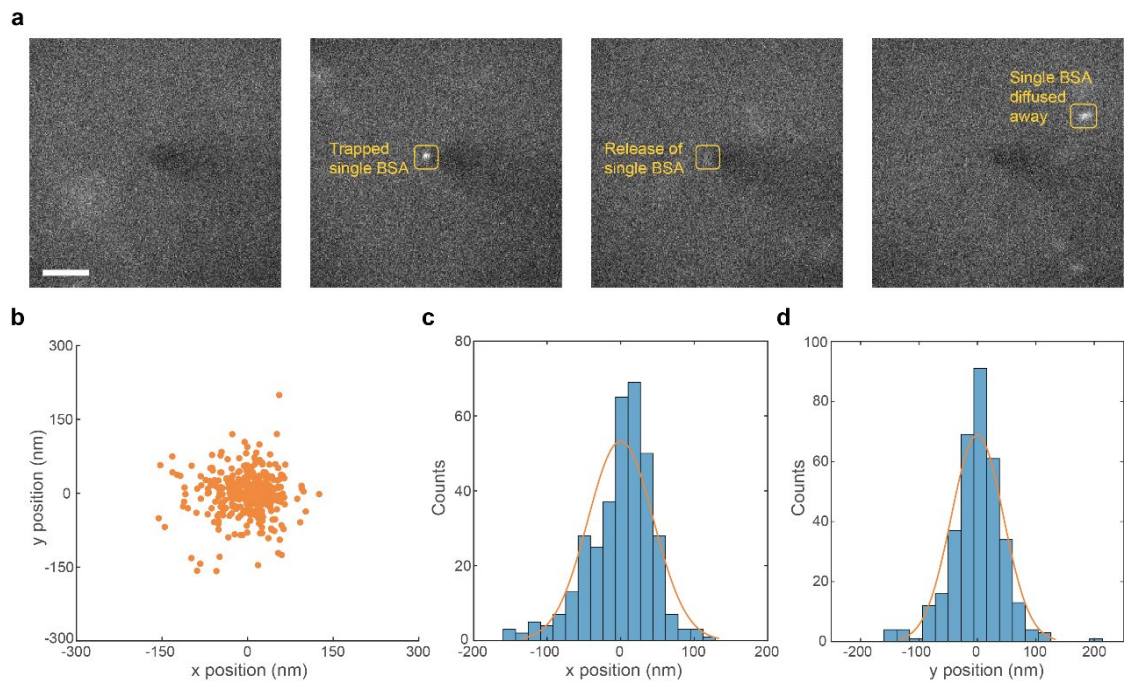

**Extended Data Fig. 3. Trapping of single BSA molecule using PET. a**, Snapshots of a trapping and release process of single BSA molecule. Scale bar: 5 μm. **b**, Transverse position of another Cy5-tagged BSA molecule captured by PET. **c**, **d**, Histograms of displacement in x direction and y direction in **b**, respectively. The orange curve corresponds to the Gaussian fitting curve.

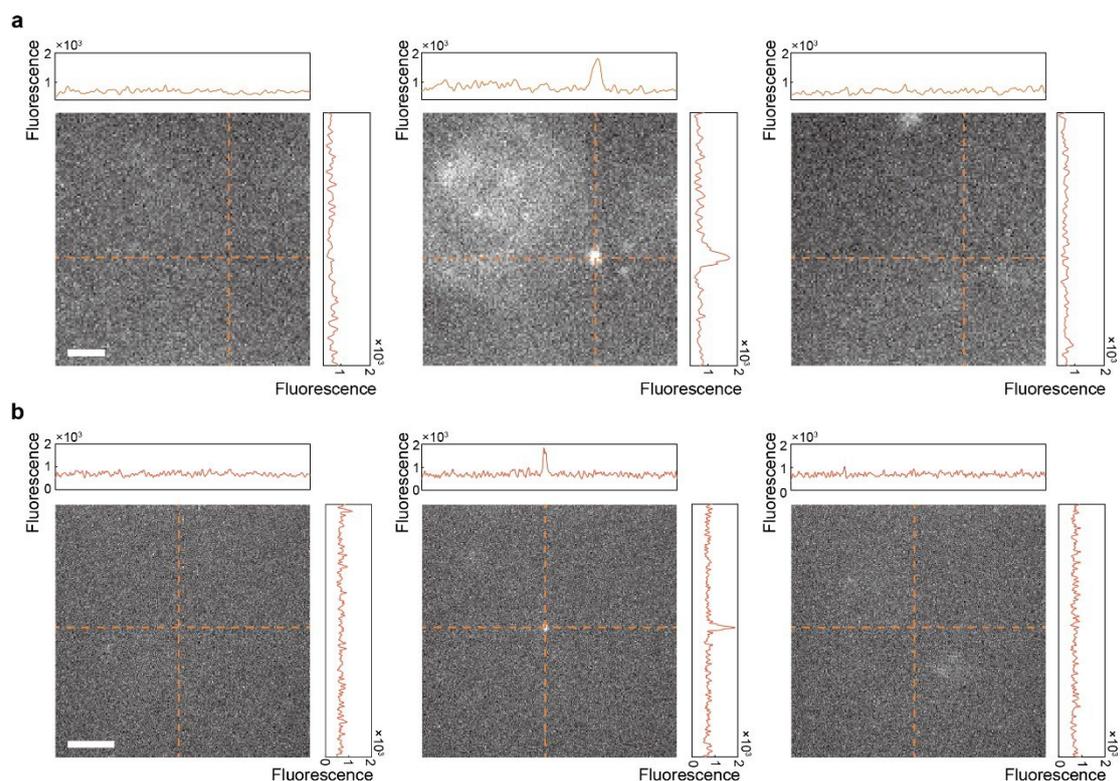

**Extended Data Fig. 4. Trapping of λDNA and 100 nm PS bead using PET. a**, Snapshots of a trapping and release process of λDNA using PET. The intersection of dashed line denoted the position of PET tip, which did not show a peak in fluorescence-position trace before trapping (left). When a.c. signals with amplitude of $V_{ac}$ = 1V and frequency $f$ = 30 MHz were applied, fluorescence-position trace showed the YOYO-1 labelled λDNA molecule was captured at the tip of PET (middle). After the a.c. signals were withdrawn, trapped λDNA molecule was released and diffused away (right). Scale bar: 2 μm. **b**, Snapshots of a trapping and release process of 100 nm PS bead using PET. Similar change in fluorescence-position trace showed the trapping and release of 100 nm PS bead before application of a.c. signal (left), during trapping (middle) and release of trapped bead (right). Scale bar: 5 μm.

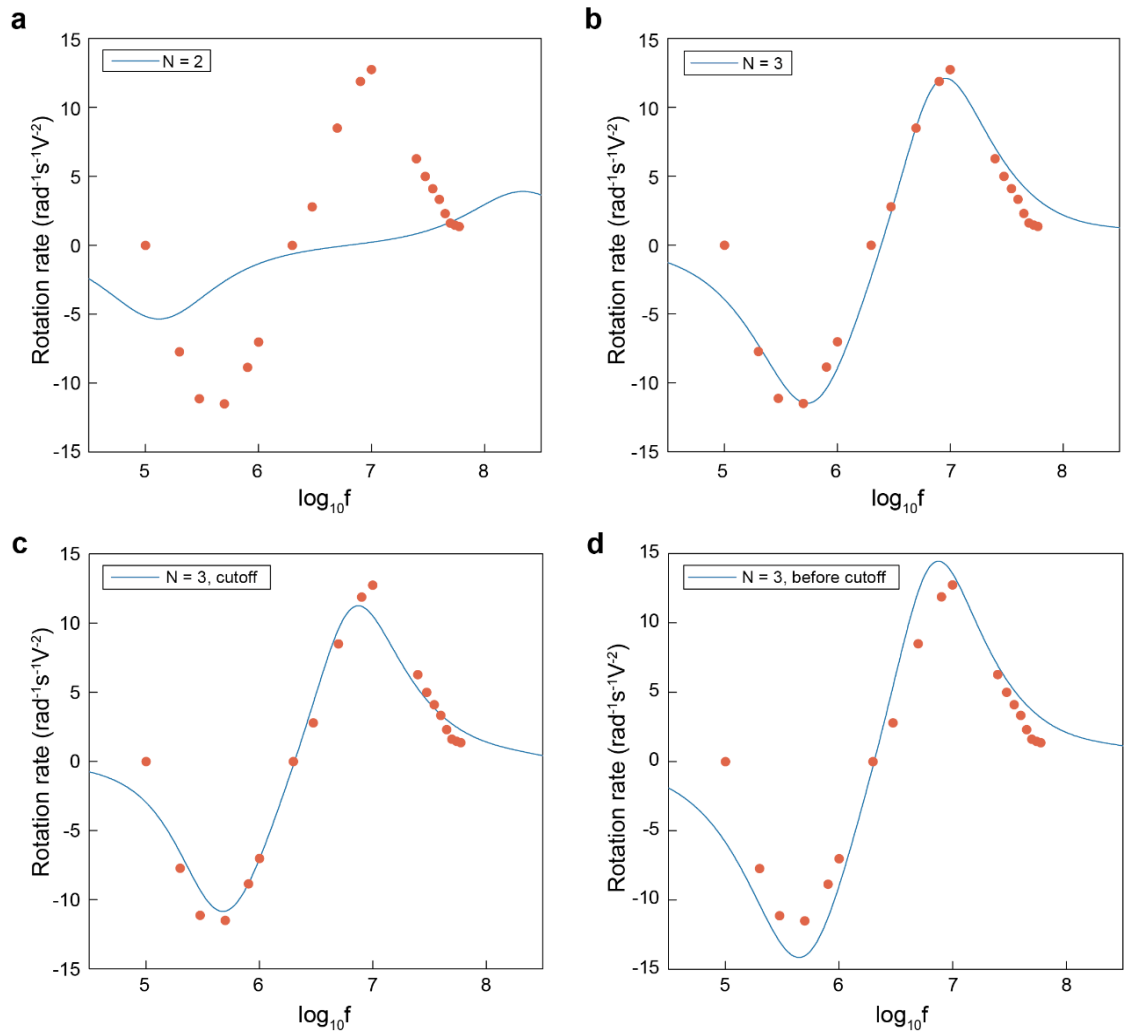

**Extended Data Fig. 5. ROT spectra of *E. coli* and fitting results.** The N = 2 model cannot capture the basic ROT spectrum features (**a**), while the N = 3 model can fit the ROT spectrum data well ($R^2$ = 0.943). The use of cutoff results can slightly improve the fitting effect, with the goodness of fit being 0.970. The introduction of the cutoff function has little impact on the fitting parameters. The fitting curve of corresponding parameters has a $R^2$ value of 0.883.

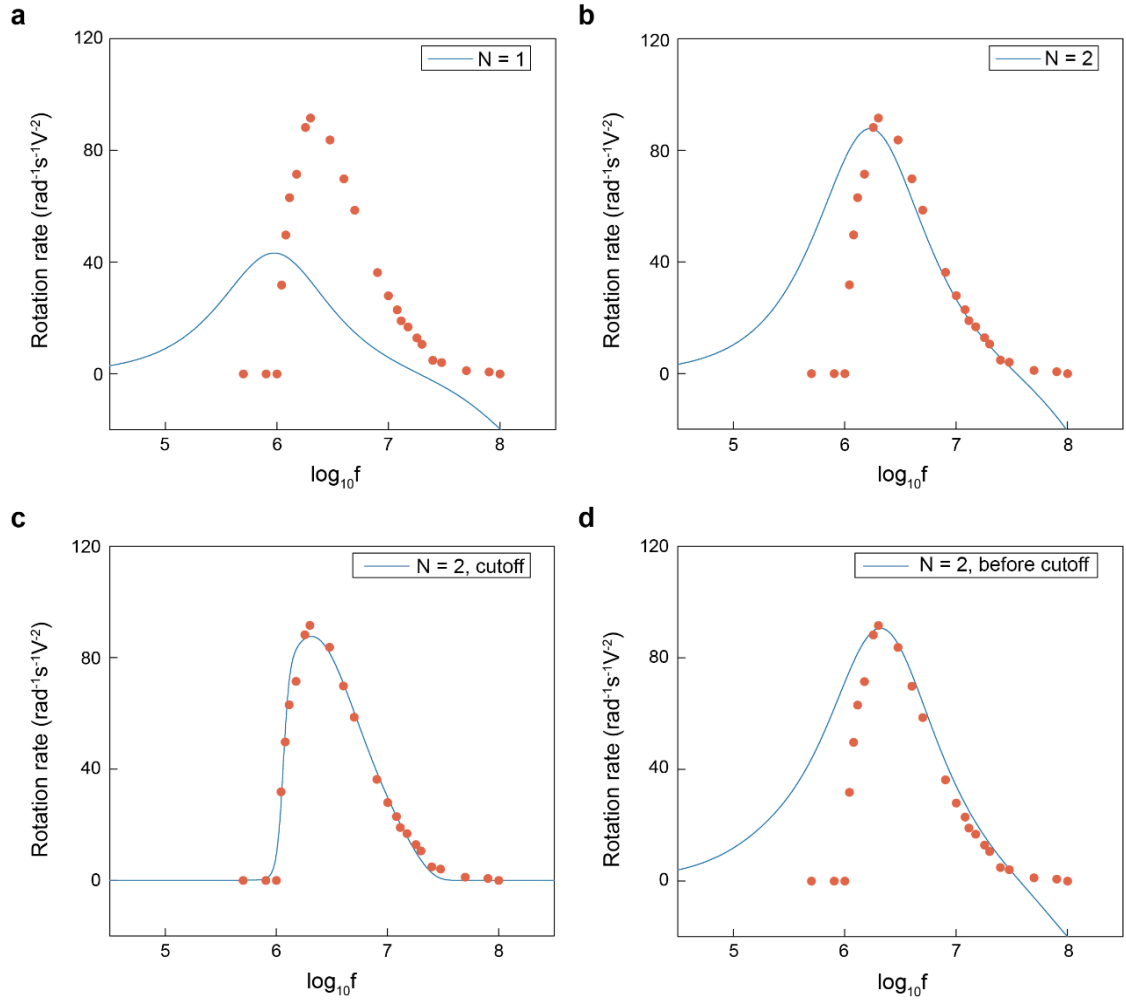

**Extended Data Fig. 6. ROT spectra of *Bifidobacterium* and fitting results.** The N = 1 model cannot capture the basic ROT spectrum features. Due to the lack of a consideration for the transition from static friction to dynamic friction, the model with N = 2 also fails to fit the observed data ($R^2$=0.232). The use of cutoff can significantly improve the goodness of fitting ($R^2$=0.983). In this system, the introduction of the cutoff function will not significantly affect the values of conductance of different shells, with the fitting curve for the corresponding parameters has a $R^2$ value of 0.356.

# Supplementary Information for
# Programmable Electric Tweezers

Yuang Chen[1], Haojing Tan[1], Jiahua Zhuang[1], Yang Xu[1], Chen Zhang[1,2]* and Jiandong Feng[1,2,3]*

[1]*Laboratory of Experimental Physical Biology, Department of Chemistry, Zhejiang University, 310027 Hangzhou, China*

[2]*The First Affiliated Hospital, School of Medicine, Zhejiang University, 310003 Hangzhou, China*

[3]*Institute of Fundamental and Transdisciplinary Research, Zhejiang University, 310058 Hangzhou, China*

*Correspondence to jiandong.feng@zju.edu.cn and chen-zhang@zju.edu.cn*

# Table of content

**Supplementary Table 4.** Comparison of single-object manipulation techniques.

**Supplementary Video 1.** Capture, release and recapture of a 1 μm bead using PET.

**Supplementary Video 2.** Trapping of single Bovine Serum albumin (BSA) molecule using PET.

**Supplementary Video 3.** Trapping of a 1 μm bead at the center of PET within the range of pDEP frequency.

**Supplementary Video 4.** Rotation of electric field in actual configuration from numerical simulation.

**Supplementary Video 5.** Electrorotation of a single *E. coli* cell.

**Supplementary Video 6.** Selective trapping and rotation of single magnetic bead with 1 μm size.

**Supplementary Note 1:** Fabrication of programmable electric tweezers.

Programmable electric tweezers (PET) were fabricated from quad-barrel capillaries by a four-step process, including pulling, carbon deposition, wet etching and copper wire connection as shown in **Supplementary Fig. 1**.

1) Quad-barrel capillaries were cleaned using a plasma cleaner (Diener electronic) for 15 minutes and then pulled via a laser puller (P-2000, Sutter Instrument) using a two-line program (Line 1: HEAT-750, FIL-3, VEL-30, DEL-160, PUL-50; Line 2: HEAT-725, FIL-3, VEL-50, DEL-140, PUL-70). It should be noted that parameters are instrument specific thus the pulling results may vary from puller to puller.

2) Quadruple carbon nanoelectrodes were generated via pyrolytic deposition of carbon at the tip of nanopipettes obtained from pulling which was reported elsewhere[1]. As shown in **Supplementary Fig. 1b**, butane was passed through the nanopipette by a Tygon tube (Saint-Gobain) and heated at the tip of nanopipette using a butane torch for 40 s under a nitrogen atmosphere to deposit carbon electrodes.

3) The tip of nanopipettes was dipped into a 10:1 buffered oxide etchant (BOE, purchased from Sigma-Aldrich) solution for different times to etch the tip to expose the carbon electrodes. It was interesting that a longer etching time yield a larger gap between opposite carbon electrodes which can further be utilized to control the characteristic size of the tip. Nanopipettes with characteristic size varying from hundreds of nanometers to several micrometers can be fabricated by changing the etching time as shown in **Supplementary Fig. 2**. It should be noted that the etching rate is related to the temperature and increase sharply with the etching time.

4) Copper wires (diameter = 0.15 mm) with one end soldered with pin header were inserted into the nanopipettes from the back end to touch the carbon electrodes. After adding a drop of glue (Ergo 5400) to fix and insulate four copper wires, the fabrication of PET was finally completed and each electrode of tweezers can be individually addressed.

Interestingly, carbon electrodes become elastic and flexible after etching process. It will bend without breaking when being pressed to the bottom glass and recover original shape when the force was withdrawn, different from the fragile glass nanopipette[2]. Owing to the loss of quartz septum, the tips of carbon electrodes will shrink and the arrangement becomes chaotic due to the surface tension when lifting out of the etching solution thus an expanding process was needed before experiment to recover the quadruple configuration. The images of a closed PET and typical PET with different characteristic sizes are acquired using scanning electron microscope (GeminiSEM300, Zeiss) as shown in **Supplementary Fig. 3**.

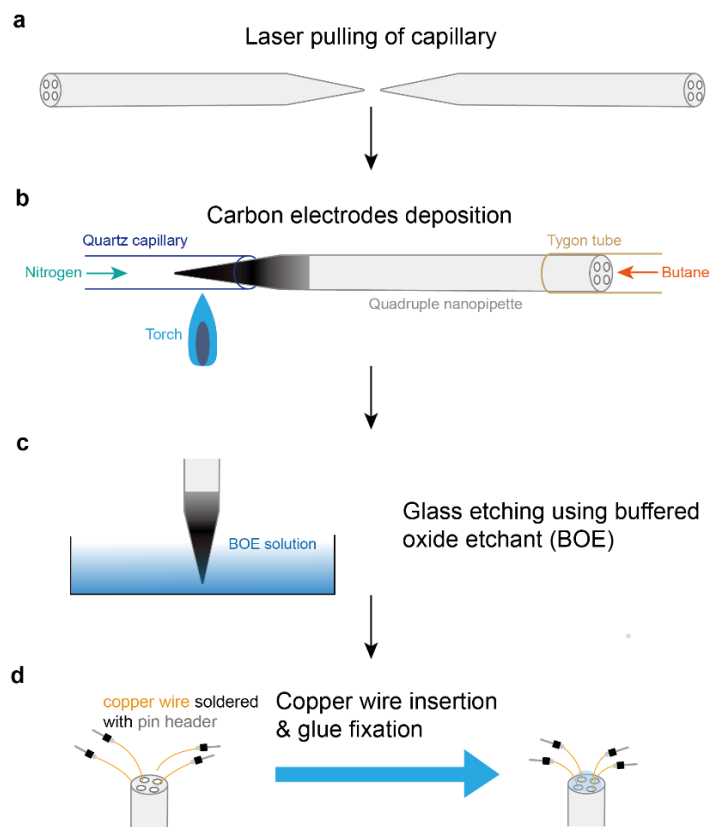

**Supplementary Fig. 1. Fabrication process of PET. a,** Quad-barreled capillaries are pulled into two nanopipettes via a laser puller. **b,** Pyrolytic deposition of carbon electrodes at the tip of nanopipettes. **c,** Wet etching of glass using BOE solution. **d,** Copper wire insertion and glue fixation to complete the fabrication of PET.

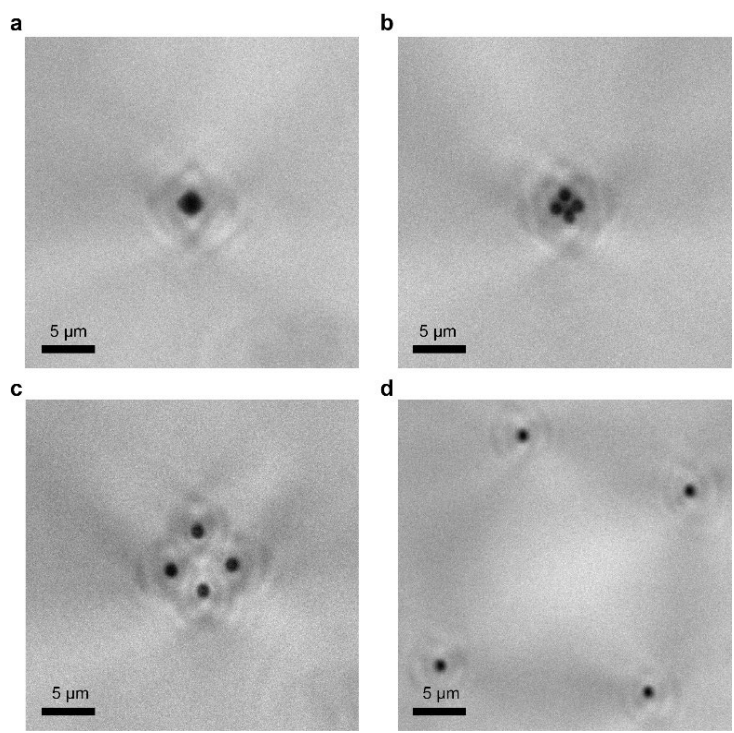

**Supplementary Fig. 2. Characteristic size of PET versus etching time.** The characteristic size increased when increasing the etching time while the size of electrodes was maintained, implying the selective etching of quartz. The characteristic size was hundreds of nanometers when etching for 0.5 h **(a)** and 1 h **(b)** and reached several micrometers when etching for 2 h **(c)** and 3 h **(d)**. PET with etching time varying from 100 min to 120 min were used in experiments which corresponds to a characteristic size between 4 μm to 8 μm owing to the influence of temperature on etching rate.

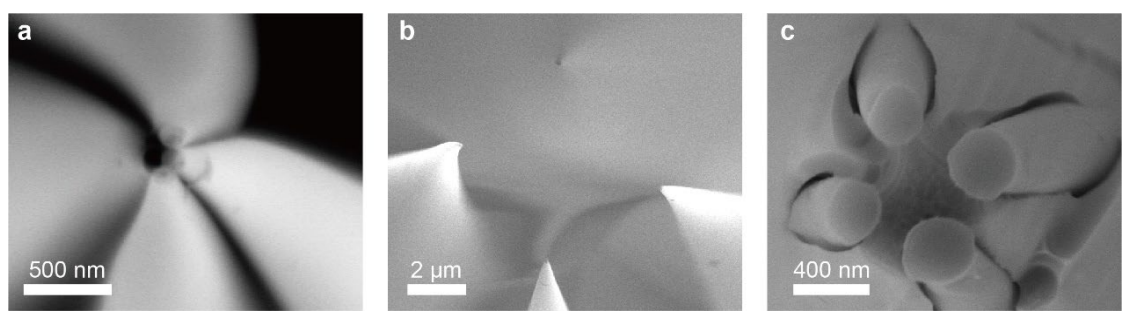

**Supplementary Fig. 3. SEM images of PET. a**, A closed tweezer that four nanoelectrodes gathered together. **b**, A typical PET that four nanoelectrodes were expanded with gap size of ~7 μm. The quartz shell was totally etched while the carbon electrode was exposed. **c**, Another PET with gap size of ~400 nm.

**Supplementary Note 2:** Principles of Paul trap and electrorotation

**Principle of Paul trap**

The trapping principle of Paul trap is based on the electrophoretic force between charged particles and inhomogeneous oscillating electrical field[3,4]. Assume a positively charged particle in quadrupole electric field that was generated from quadruple electrodes with voltages in the form of $\pm(U_{\text{dc}} - V_{\text{ac}} \cos \Omega t)$ and adjacent electrodes have opposite voltages. As shown in **Supplementary Fig. 4a**, the potential distribution had a saddle-like shape thus the positively charged particle will be repelled into the center of tweezer in x direction while pulled away from the center alone the y direction at t = 0. After half of a period, the polarity reversed thus the particle will experience the opposite force (**Supplementary Fig. 4b**). If the electric field oscillate at an appropriate frequency, the particle will be dynamically stuck in this back-and-forth motion. In other word, the charged particle was captured in the pseudopotential well in the center of tweezer as shown in **Supplementary Fig. 4c**. It should be noticed that negatively charge particle also experience a similar process and enough charge was required for Paul trap since it originated from the electrophoresis force.

**Principle of electrorotation**

Different from Paul trap that require net charge, electrorotation (ROT) need polarization of the object. When a polarizable particle is located in the rotating electric field, it has a dipole and should rotate synchronously with the rotating electric field. However, if the frequency of field is too high for the relaxation of the particle to achieve synchronization, the particle will be exerted by a torque originated from the phase delay:

$$\Gamma_{\text{ROT}} = -4\pi r_{eff}^3 \varepsilon_m Im[K(\omega)] \boldsymbol{E}^2 \quad (1)$$

where $\varepsilon_m$ is the permittivity of medium, $r_{eff}$ is the effective radius of particle, $\boldsymbol{E}^2$ is the square of electrical field, and $Im[K(\omega)]$ is the imaginary part of Clausius-Mossotti (CM) factor in an alternating current (a.c.) electric field of angular frequency $\omega$, which is depended on the complex permittivity of medium $\varepsilon_m^*$ and the complex permittivity of particle $\varepsilon_p^*$ by

$$K(\omega) = \frac{\varepsilon_p^* - \varepsilon_m^*}{\varepsilon_p^* + 2\varepsilon_m^*} \quad (2)$$

The complex permittivity for an isotropic homogeneous dielectric is given by

$$\varepsilon^* = \varepsilon - j\frac{\sigma}{\omega} \quad (3)$$

where $\varepsilon$, $\sigma$ is the permittivity and conductivity of the dielectric, respectively, and $j$ is the imaginary unit.

Since the electrorotation torque in equation (1) balances with the sum of hydrodynamic resistance torque and interfacial friction torque ($\Gamma_{ROT} = \Gamma_H + \Gamma_{friction}$), it can be written as $\Gamma_{ROT} = \alpha w$ owing to the linear relationship between the friction as well as hydrodynamic resistance and the rotation speed (**Supplementary Note 7**), where $\alpha$ is a coefficient, $w$ is the angular velocity. Further, the electric field $\boldsymbol{E}$ is generated by the voltage applied on the electrodes ($\boldsymbol{E} \propto V_{ROT}$), yielding the angular velocity $w \propto V_{ROT}^2$, which was experimentally observed.

Effective radius was introduced to model particles which were not homogeneous sphere by involving the conductivity and permittivity of different parts (e.g. cell membrane and cytoplasm) as reported by former works[5,6]. Utilizing the cell modelling, dielectric parameters can be obtained by fitting the experimental ROT spectrum (plot of rotation rate versus field frequency) with the theoretical spectrum.

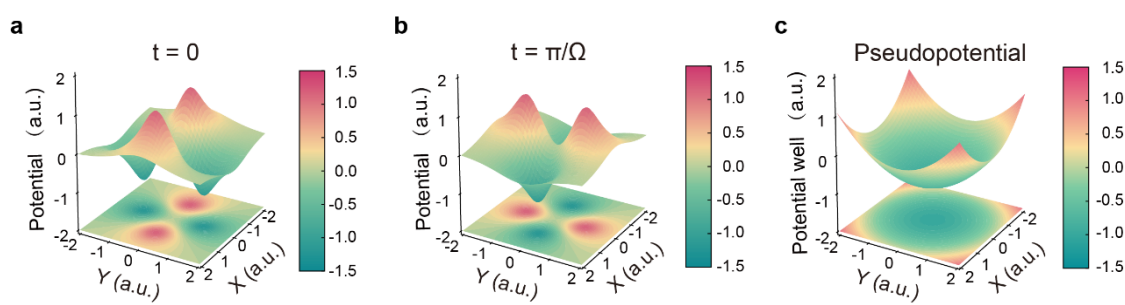

**Supplementary Fig. 4. Potential distribution and pseudopotential of Paul trap. a**, Potential distribution of Paul trap when t = 0. **b**, Potential distribution of Paul trap when t = π/Ω. **c**, Pseudopotential of Paul trap generated by the inhomogeneous oscillating electrical field.

**Supplementary Note 3:** Numerical simulation of PET for electric field distribution of Paul trap and further demonstration of potential programmability.

**Numerical simulation of PET for electric field distribution of Paul trap**

COMSOL Multiphysics was used to perform finite-element method to model electric field of PET. A three-dimensional model was established to mimic the configuration of quadruple electrical tweezer in experiments as shown in **Supplementary Fig. 5**. According to results obtained by SEM, the glass shell was totally etched and all electrodes were exposed at the very tip of the tweezer. Thus, we introduce the first 8 μm of the tip which was all consisted of pyrolytic carbon into the model. The parameters of geometry were specified in **Supplementary Table 1**. Briefly, the vertically aligned quadruple electrical tweezer with length of $L$ was surrounded by the square solution with a side length of $SL$. Each carbon electrodes had the same tip diameter $d$ and same bottom diameter $D$. The distance between opposite electrodes (denoted as characteristic size of quadruple electrical tweezer) was $2R_0$. Parameters of materials used in simulations were specified in the **Supplementary Table 2**.

We used *Electric Currents* as physical field to solve time dependent results of simulation to obtain the electric field distribution of Paul trap. All boundaries of quadruple PET are applied with electric potential with the form of $\pm V_{ac}\cos(2\pi f t)$ where $V_{ac} = 2V, f = 1MHz$, adjacent electrodes have the same potential and opposite electrodes have opposite potentials. Mesh was controlled by the physical field with an extra fine resolution.

**Supplementary Table 1. Geometry parameters for numerical simulations**

| Geometry parameter | Definition | Values |
|---|---|---|
| $2R_0$ | Characteristic size of PET | 6 μm |
| d | Tip diameter of tweezer in model | 0.2 μm |
| D | Bottom diameter of tweezer in model | 2 μm |
| L | Length of carbon electrodes in model | 8 μm |
| SL | Side length of solution | 10 μm |

**Supplementary Table 2. Parameters of material used in numerical simulations.**

| Material | Relative permittivity (ε) | Electrical conductivity (σ) [S·m$^{-1}$] |
|---|---|---|
| Water | 78 | $5.5 \times 10^{-6}$ |
| Carbon | 1.8 | $1.2 \times 10^3$ |

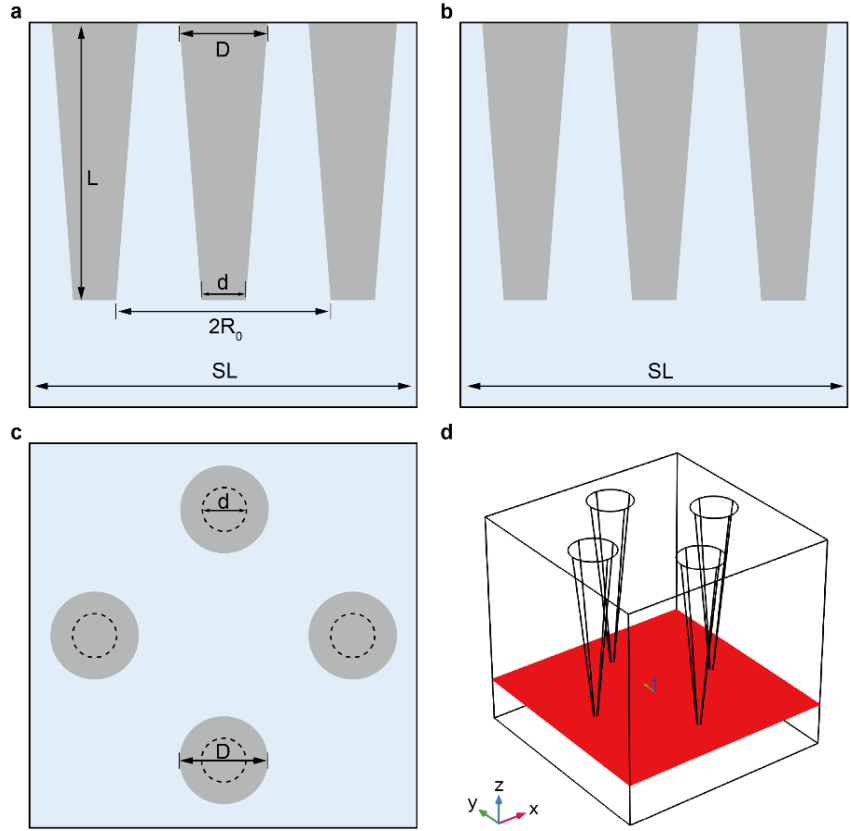

**Supplementary Fig. 5. Geometry model for PET used in numerical simulations. a-c,** Front view, side view and top view of geometry model, respectively. Diagrams are not to scale. **d,** Axonometric view of geometry model. Red cut plane was introduced to plot electric field distribution at the tip plane.

**Numerical simulation of PET for further demonstration of potential programmability.**

To further demonstrate the local potential programmability of PET, A two-dimensional model was established as shown in **Extended data Fig. 1**. As a proof of concept, we designed a 6×6 electrode array with each electrode possessing a radius $r$ = 100 nm and a spacing $a$ = 2 μm, which was placed in water with a liquid length $LL$ = 200 μm. Since each nanoelectrode of PET is addressable, programmable signal can be applied to the array to modify the local potential. Here we design the potential on each electrode as shown in **Extended data Fig. 1** to create potential distribution in the shape 'LE' and shape 'PE' since abbreviation of our lab is 'LEPB'. Other parameters and settings used in the simulation are identical to those in the previous simulation for distribution of Paul trap.

The potential distribution in the shape of 'LE' and 'PB' is clearly shown in **Supplementary Fig. 6**. Therefore, spatiotemporally editing the potential landscape to create pattern or achieve certain function via PET is feasible. Furthermore, we extract the electric field strength out along a line across the center of 'PB' and plot the electric field strength as a function of the x-position to investigate the range influenced by PET, as shown in **Extended data Fig. 2**. Considering a decrease to 1% of the peak strength as indicating no influence, we can determine that the range influenced by PET is about 25 μm, demonstrating the localization of the electric field generated from PET.

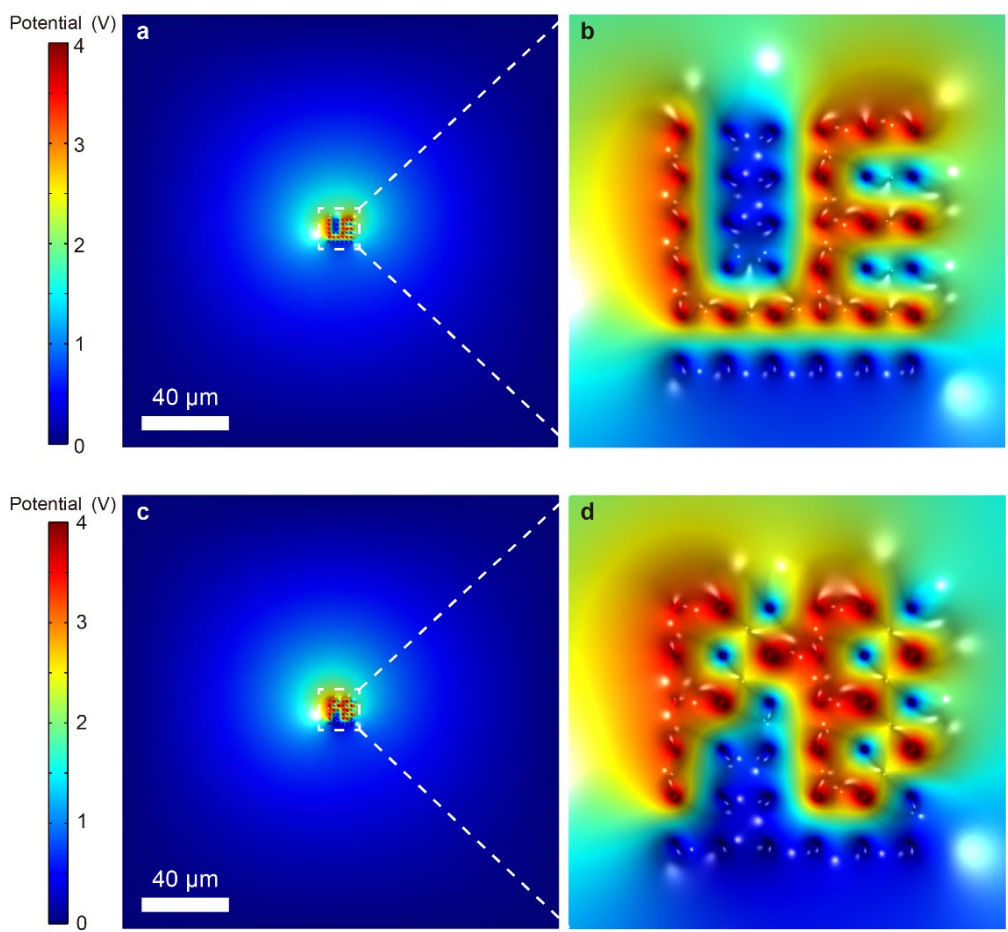

**Supplementary Fig. 6. Programmed potential distribution in the shape of 'LE' and 'PB' via PET. a, b,** Potential distribution from the simulation results for 'LE' pattern where **b** is the magnified view of the central region in **a**. The potential distribution in the shape of 'LE' is clearly shown. **c, d,** Potential distribution from the simulation results for 'PB' pattern where **d** is the magnified view of the central region in **c**. The potential distribution in the shape of 'PB' is clearly shown.

**Supplementary Note 4:** Trapping of single biomolecules and nanoparticles using PET.

Single biomolecules including bovine serum albumin (BSA) and λDNA can also be captured by PET. As shown in **Extended Data Fig. 3** and **Supplementary Video 2**, Single Cy5-tagged BSA can be captured and released by a PET made from an adjusted fabrication process (modified pulling parameter and dry etching process for 10 min using $CF_4$ to obtain PET in nanoscale). The trapping was achieved by applying a.c. signals with voltage $V_{ac}$ = 1.5V at frequency $f$ = 30MHz. In addition, λDNA (labelled with YOYO-1) and 100 nm PS bead were selected as another trapping targets as shown in **Extended Data Fig. 4**, demonstrating the wide range of trapping target from microscale entities to biomolecules and particles in nanoscale for PET.

**Supplementary Note 5:** Verification of trapping mechanism in PS bead trapping

Dielectrophoresis (DEP) was inevitable when applying a.c. voltage to quadruple electrodes tweezers. For a homogeneous dielectric particle in aqueous solution, the average DEP force ($\boldsymbol{F}_{DEP}$) was given by[7]

$$\boldsymbol{F}_{DEP} = 2\pi\varepsilon_m r^3 Re[K(\omega)]\nabla E^2 \qquad (4)$$

where $\varepsilon_m$ is the permittivity of medium, $r$ is the radius of particle, $Re[K(\omega)]$ is the real part of Clausius-Mossotti (CM) factor in an a.c. field of angular frequency $\omega$, and $\nabla E^2$ is the gradient of the square of electrical field.

If $Re[K(\omega)] > 0$, the dielectric particle will be attracted to the location with a stronger electric field, which is referred as positive DEP (pDEP). In contrast, if $Re[K(\omega)] < 0$, the particle will be pushed away from high field regions, which is referred as negative DEP (nDEP).

According to equation (2) and (3), the behavior of PS bead influenced by DEP is governed by properties of particle and solution conditions. The conductivity of beads with different size can be evaluated by employing the Maxwell-Wagner-O'Konski (MWO) model[8], which is given by

$$\sigma_p = \sigma_b + 2K_s/r \qquad (5)$$

where $\sigma_b$ is the bulk conductivity and $K_s$ is the surface conductance of particle. Using parameters shown in **Supplementary Table 3**, the real part of CM factor can be calculated and the curve of $Re[K(\omega)]$ versus frequency for 1 μm PS bead was shown in **Supplementary Fig. 7**. According to the curve, the crossover frequency ($Re[K(\omega)] = 0$) is 1.89 MHz.

**Supplementary Table 3. Parameters for CM factor calculation.**

| Parameters | Values |
|---|---|
| Permittivity of medium ($\varepsilon_m$) | 78.5 $\varepsilon_0$ |
| Conductivity of medium ($\sigma_m$) | 4 μS·cm$^{-1}$ (Measured) |
| Radius of PS bead ($r$) | 500 nm |
| Surface conductance of PS bead ($K_s$) | 2.85 nS |
| Permittivity of PS bead ($\varepsilon_p$) | 2.55 $\varepsilon_0$ |
| Bulk conductivity of PS bead ($\sigma_b$) | 10$^{-14}$ S·m$^{-1}$ |

For PET, the electric field distribution was numerically simulated via COMSOL and shown in **Supplementary Fig. 8**. The center of quadruple electrical tweezer had lower electric field strength compared with edges of electrodes. Therefore, pDEP will force the bead to electrodes while nDEP forces the bead to the center or away from the tweezer.

As shown in **Supplementary Video 3**, the trapping phenomenon of 1 μm PS bead at the center of quadruple electrical tweezer was observed at different frequencies which range from pDEP region (<crossover frequency) to nDEP region while other beads out of the tweezer was attracted to the electrodes, showing the characteristic of pDEP. When the frequency is located in pDEP region, the bead should be pushed away from the center but the trapping at the center was observed during the experiment, implying Paul trap rather than DEP effect takes the dominant role in trapping. Therefore, we attribute the trapping mechanism to Paul trap rather than DEP effect.

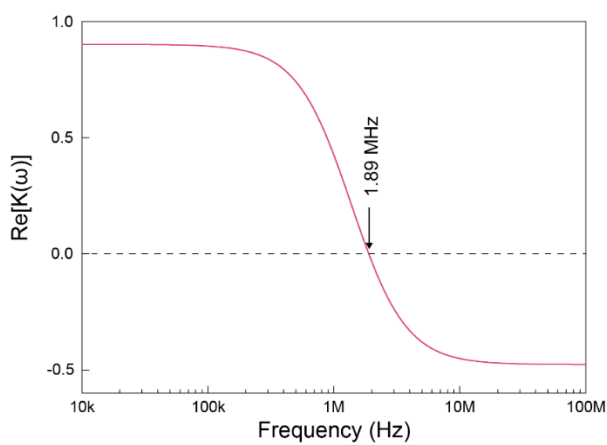

**Supplementary Fig. 7. Calculated real part of CM factor for 1 μm PS bead.** Curve of $Re[K(\omega)]$ versus frequency for 1 μm PS bead shows the crossover frequency is 1.89 MHz.

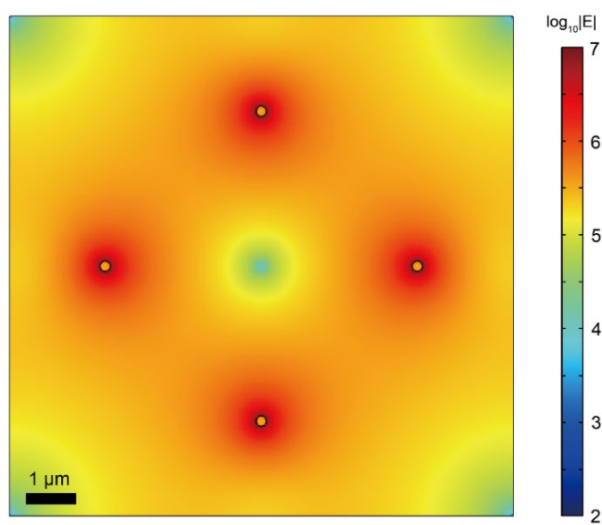

**Supplementary Fig. 8. Mapping of $\log_{10}|E|$ for PET via numerical simulation.** The center of PET had a relative lower electric field strength while edges of electrodes had the highest strength of electric field.

**Supplementary Note 6:** Effective specific charge measurement via Paul trap in aqueous solution.

For an ideal planar quadrupole electric field generated from the applied voltages, the potential distribution varied with time in x-y plane was assumed as[4]

$$\varphi(x, y, t) = (U_{dc} - V_{ac} \cos\Omega t) \frac{x^2 - y^2}{2R_0^2} \quad (6)$$

For a homogeneous particle with mass $M$, radius $r$ and effective net charge $Q$ in solution, we exclude the effects of buoyancy, electrothermal flow and a.c. electro-osmotic flow on its motion since their effect is negligible for the micro volume of targets, low heating effect of our devices and relatively high frequency used in experiments (>100 kHz) respectively thus the motion of this particle driven by electric field force, Stokes drag force and Brownian motion force can be described as

$$M \frac{d^2 \vec{r}}{dt^2} = -\xi \frac{d\vec{r}}{dt} + (-\nabla\varphi)Q + \vec{N}(t) \quad (7)$$

where $\vec{r}$ is the object position vector in x-y plane, $\xi$ is the Stokes' drag coefficient as approximated by $\xi = 6\pi\eta r$, $\eta$ is the dynamic viscosity of solution, $\vec{N}(t)$ is a random force caused by thermal fluctuations. Rewrite the equation (7) into two Langevin equations in x and y direction by introducing dimensionless scaled time $\tau = \Omega t/2$, scaled dc voltage $a = 4QU_{dc}/MR_0^2\Omega^2$, scaled ac voltage $q = 2QV_{ac}/MR_0^2\Omega^2$, scaled damping coefficient $b = 2\xi/M\Omega$ and the scaled thermal fluctuation force $g(\tau)$,

$$\frac{d^2 x}{d\tau^2} + b \frac{dx}{d\tau} + (a - 2q\cos 2\tau)x = g(\tau) \quad (8)$$

$$\frac{d^2 x}{d\tau^2} + b \frac{dx}{d\tau} + (a - 2q\cos 2\tau)x = g(\tau) \quad (9)$$

Indeed, the geometry in our experiments was not an ideal two-dimensional Paul trap structure owing to the cone shape of quadruple electrical tweezers. Therefore, a correction factor $\Gamma$ is introduced to deal with this bias as well as the higher-order components in nonideal quadruple electrical field[9], thus the parameter $a$ and $q$ can be written as

$$a = \frac{4QU_{dc}}{\Gamma MR_0^2\Omega^2} \text{ and } q = \frac{4QV_{ac}}{\Gamma MR_0^2\Omega^2} \quad (10)$$

Different from the Paul trap in vacuum (b = 0 and $g(\tau)$), the well-known stability diagram will experience shift and extension if the Strokes drag force is present while Brownian motion and DEP effect does not affect stability boundaries where $q > 0.1$ in our experiment[9,10].

By carefully adjusting $U_{\text{dc}}$ and $V_{\text{ac}}$ at a fix frequency of 3 MHz to search for the verge where the motion of trapped bead was no longer stable, we can obtain numerous boundary points. Utilizing the fitting parameter $\frac{Q}{\Gamma M}$ and equation (10), boundary points can be changed and plotted on the *a-q* diagram, which can be compared with theoretically calculated results.

**Supplementary Note 7:** Biological samples preparation.

**Cell culture**

The E. coli (BL21) cells were obtained from GenScript Biotech and the Bifidobacterium cells (No.21710) were obtained from China center of industrial culture collection (CICC, Beijing). All bacteria were stored in glycerol stock at -80 °C. Before use, cells of E. coli were recovered from frozen condition and incubated in LB medium (Aladdin Biochemical Technology) overnight at 37 °C. The next day, the pre-culture was diluted 1000-fold into fresh LB medium and incubated at 37 °C with shaking by 200 revolutions per minute (rpm) to reach the logarithmic phase (OD600nm = 0.4 - 0.5) for further use. Cells of Bifidobacterium were recovered from frozen condition and incubated in RCM medium overnight at 37°C in the absence of oxygen. The next day, the pre-culture was diluted 1000-fold into fresh RCM medium (Solarbio Science & Technology) and incubated at 37 °C with shaking (200 rpm) to reach the logarithmic phase (OD600nm = 0.4 - 0.5) for further use.

**Torsion-constrained DNA preparation.**

Firstly, A multiple-biotin-labeled short fragment was PCR amplified from lambda DNA using primers (primer sequences: 5'-GCTTGGCTCTGCTAACACG-TTGCTCATAGGAG-3' and 5'-CAGCTACAGTCAGAATTTATTGAAGCAA-3', purchased from Sangon Biotech) together with 30% biotin-11-dUTP (NU-803-BIOX-S, Jena Bioscience). Secondly, a multiple-digoxigenin-labeled short fragment was PCR amplified from lambda DNA (Thermo Fisher) using primers (primer sequences: 5'-CCCTAAGACCTTTAATATATCGCCAAATAC-3' and 5'-AATTTAGCCC-TTCAATCGCCAGAGAAATCTAC-3', purchased from Sangon Biotech) together with 30% digoxigenin-11-dUTP (NU-803-DIGX-S, Jena Bioscience). Finally, the torsion-constrained DNA construct was PCR amplified from lambda DNA using the two short DNA fragments as megaprimers.

**Supplementary Note 8:** Actual electric field configuration.

In experiments, PET was connected to a function generator (AFG31000, Tektronix) which has two channels with phase difference of 90°. As shown in **Supplementary Fig. 9a**, electrode 1 (E1) and electrode 3 (E3) are connected to channel 1 while electrode 2 (E2) and electrode 4 (E4) are connected to channel 2 to generate a rotating electric field. However, owing to the limitation that function generator outputs with a common ground, the actual phase situation is shown in **Supplementary Fig. 9b**, which is different from an ideal rotating electric field.

Nevertheless, the rotation was still observed in experiment and we also carried out numerical simulation for this situation. Based on the model in **Supplementary Note 3**, signals with the form of $V_{\text{ROT}}\sin(2\pi ft + \varphi)$ are applied to corresponding electrodes where $V_{\text{ROT}} = 1$ V, $f = 1$ MHz and $\varphi$ is the phase of electrode. Although E3 and E4 bear the same potential, the rotation of electric field in the region within PET is still observed and almost uniform as shown in **Supplementary Video 4**.

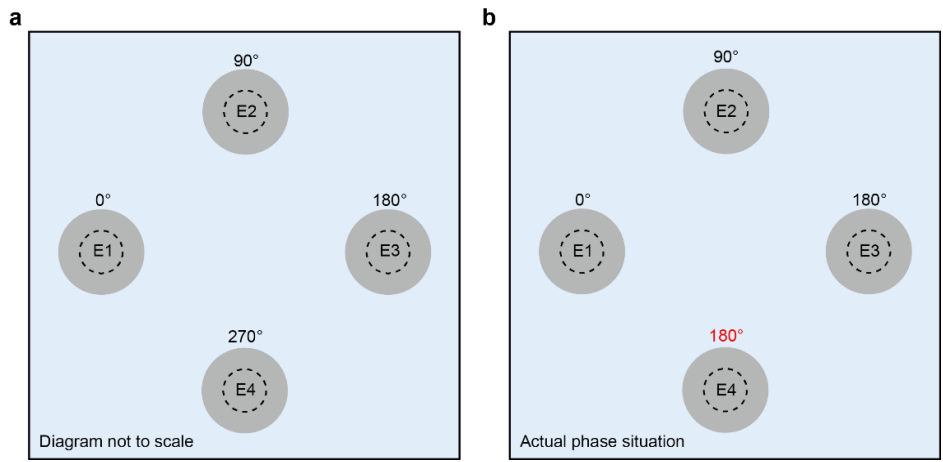

**Supplementary Fig. 9. Ideal phase and actual phase situation in ROT experiments. (a)** Ideal configuration in ROT experiments. **(b)** Actual configuration in ROT experiments owing to the limitation of function generator.

**Supplementary Note 9:** Fitting of electrorotation spectra for partially fixed bacteria.

**Dielectric model for microalgae cell:**

We calculated the effective complex permittivity of *Bifidobacterium* and *E. coli* using two-shell and three-shell dielectric models, respectively. For the multi-shell dielectric model, its numerical values were calculated via a smeared-out sphere approach as shown in below:

$$S(a,b,c,d) = b\left[\left(\frac{d}{c}\right)^3 + 2\frac{a-b}{a+2b}\right] \bigg/ \left[\left(\frac{d}{c}\right)^3 - \frac{a-b}{a+2b}\right] = b/f\left[\left(\frac{d}{c}\right)^3, f(a,b)\right]$$

where f is the Clausius-Mossotti factor:

$$f(a,b) = \frac{a-b}{a+2b}$$

For a double-layer model (N=2), the effective complex permittivity is:

$$\varepsilon_{p,eff}^* = S(S(\varepsilon_1^*, \varepsilon_2^*, R_1, R_2), \varepsilon_3^*, R_2, R_3)$$

For a three-layer model (N=3), the effective complex permittivity is:

$$\varepsilon_{p,eff}^* = S(S(S(\varepsilon_1^*, \varepsilon_2^*, R_1, R_2), \varepsilon_3^*, R_2, R_3), \varepsilon_4^*, R_3, R_4)$$

Here, $\varepsilon_i^*$ represents the complex permittivity of the i-th shell, which is defined as $\varepsilon_i - \frac{\sigma_i}{\omega}j$. The $\varepsilon_i$ and $\sigma_i$ is the permittivity and conductance of i-th shell, respectively. $R_i$ is the thickness of i-th shell. $\omega$ is the frequency of external electric field and $j$ is the imaginary unit.

**Hydrodynamic friction and surface effect:**

In experiments, the bacteria attached to the substrate experiences hydrodynamic resistance and substrate friction when rotated by the ROT force. The hydrodynamic resistance follows a linear form:

$$f_H = 6\pi R u \eta$$

Where $R$ is bacterial radius, $u$ is the linear velocity of rotation, $\eta$ is the dynamic viscosity.

The form of interfacial friction is complex and goes beyond the scope of our study. Here, we only consider a simple case that the bacteria will only begin rotate when the rotational torque exceeds a threshold. Thus, we use a phenomenological approach to fit the data obtained from experimental observations. Two cutoff functions, $\phi_L(\omega, \omega_{L0}, B_L)$ and $\phi_R(\omega, \omega_{R0}, B_R)$, were introduced to fit the ROT spectra, where $\omega_0$ corresponds to the critical frequency and B controlling the steepness of the step function.

$$\phi_L(\omega, \omega_{L0}, B_L) = \left[1 + \tanh\left(\frac{\omega - \omega_{L0}}{B_L}\right)\right]/2$$

$$\phi_R(\omega, \omega_{R0}, B') = \left[1 - \tanh\left(\frac{\omega - \omega_{R0}}{B_R}\right)\right]/2$$

When the applied electric field frequency is within certain intervals, the torque applied is less than the maximum static friction torque, preventing overall rotation. In the remaining intervals, the friction on bacteria is linearly depending on the speed, allowing fitting of experimental curves as follows:

$$\Omega_{fit} = A\phi_L(\omega)\phi_R(\omega)\text{Im}[f(\varepsilon^*_{p,eff}, \varepsilon^*_m)]$$

$$L(\omega, \text{popt}) = \Sigma_i \left|\Omega_{exp}(\omega_i) - \Omega_{fit}(\omega_i, \varepsilon^*_{p,eff}, \varepsilon^*_m)\right|^2$$

where $\Omega_{exp}(\omega)$ is the experimentally measured rotation rate. Reasonable estimates of bacterial parameters can be obtained via optimizing the objective function L.

**Results:**

We applied the above methods to process the ROT spectra data of *E. coli* and *Bifidobacterium* on substrates (shown in **Extended Data Fig. 5** and **Extended Data Fig. 6**). The effect of the number of shells in the model and the cutoff function on fitting is demonstrated sequentially.

As the number of shells in the model increases, a more ideal fit can be achieved when the number of layers equals the actual cell structure of the bacteria. The introduction of the cutoff function can better fit the phenomenon of bacteria stopping rotation due to interfacial friction.

**Supplementary Note 10:** Extraction of Z-position changes of bead during DNA supercoiling measurements

The relative position change of trapped bead in Z direction in measurements was extracted by establishing a standard curve to obtain the relationship between the relative position of bead and intensity in the region of interest (ROI). Since the bead was tightly captured when only the Paul trap signals were applied to the PET, we can obtain images of bead at different relative positions to focal planes by utilizing the motorized focal plane control of microscope (IX83, Olympus) as shown in **Supplementary Fig. 10a**. The intensity of bead within ROI changed with relative position of bead thus we can establish a standard curve as shown in **Supplementary Fig. 10b**. A polynomial fitting with $R^2 = 0.9964$ described the relationship between the relative position of bead and the intensity of ROI. When the electrorotation signals were superposed on the PET, DNA supercoiling occurred thus the relative position of bead decreased. Therefore, the Z-position changes of bead can be determined by measuring the intensity of ROI during the experiment.

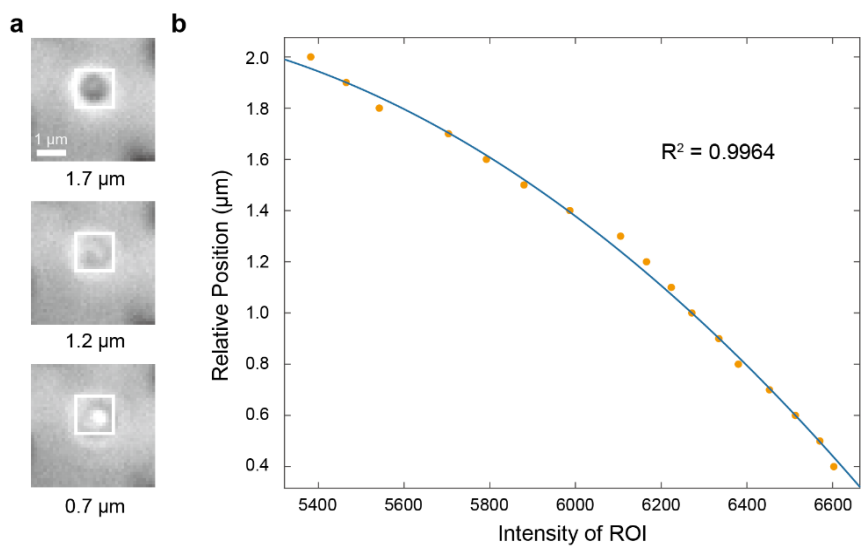

**Supplementary Fig. 10. Establishment of standard curve for extraction of Z-position changes of bead during DNA supercoiling measurements. a**, Images of a trapped bead at typical relative position by changing the focal plane. **b**, Standard curve from a polynomial fitting correlates the relative position of bead and intensity of ROI with $R^2 = 0.9964$.

**Supplementary Table 4. Comparison of single-object manipulation techniques**

| Approaches | Degree of freedom | Interacting mechanism | Functions |
|---|---|---|---|
| Optical tweezers[11] | $2^1$ | Optical force | Trapping |
| Magnetic tweezers[12] | $2^1$ | Magnetic force | Trapping |
| DEP tweezers[13] | $2^2$ | Dielectrophoretic force | Trapping, Movement |
| PET (This work) | $2^4 \to 2^n$ | Programmed electric field | Trapping, movement, rotation, multi-stepped function sequence, etc. |

## Supplementary Videos

**Supplementary Video 1. Capture, release and recapture of a 1 μm bead using PET.** This video demonstrates the capture, release and recapture process of a 1 μm bead using PET.

**Supplementary Video 2. Trapping of single Bovine Serum albumin (BSA) molecule using PET.** This video demonstrates a trapping and release process of single BSA molecule using PET.

**Supplementary Video 3. Trapping of a 1 μm bead at the center of PET within the range of pDEP frequency.** This video demonstrates the trapping of a 1 μm bead at the center of PET within the range of the pDEP frequency, confirming that the Paul trap plays a dominant role in the trapping process.

**Supplementary Video 4. Rotation of electric field in actual configuratioin from numerical simulation.** This video demonstrates the rotation of the electric field based on numerical simulation, where the arrows represent the logarithmic electric field vector and the colormap indicates the potential. Despite the limitations of the function generator, the counterclockwise rotation of the electric field is nearly uniform.

**Supplementary Video 5. Electrorotation of a single *E. coli* cell.** This video demonstrates the electrorotation of single *E. coli* at a speed of ~1000 rpm. The video is played at 0.1× speed.

**Supplementary Video 6. Selective trapping and rotation of single magnetic bead with 1 μm size.** This video demonstrates the selective and multifunctional manipulation features of PET. The magnetic bead with attachment of torsional-constrained DNA was first trapped and lifted, thus other beads were out of focus (below the focal plane). When the ROT signal was superposed, the decrease of trapped bead in Z position and a CCW rotation were observed, indicating the formation of DNA supercoiling. After removing the ROT signal, a CW rotation as well as an increase in Z position of the bead was observed, implying the relaxation of DNA supercoiling. During the process, other beads in the same field of view were not influenced.

## Supplementary References